\documentclass[12pt]{article}
\usepackage{graphicx,a4}   
\begin{document}

\newcommand{\preprintno}[1]
{{\normalsize\begin{flushright}#1\end{flushright}}}

\title{\preprintno{{\bf SUSX-TH-00-016}\\ hep-th/0010142}Topological Objects in 5D Maxwell Einstein Supergravity}
\author{Malcolm Fairbairn\thanks{E-mail address: mdsf@star.cpes.sussex.ac.uk}\\
{\em Centre for Theoretical Physics/Astronomy Centre}\\
{\em University of Sussex,} \\
        {\em Brighton BN1 9QH, U.K.}}
\date{12th October 2000}

\maketitle
\begin{abstract}
In this letter is shown that it is possible to obtain scalar hypersurfaces in 5D N=2 SUGRA where the allowed regions with positive definite scalar metric have a non-trivial topology.  This situation may aid in the construction of domain wall solutions which confine gravity to 4 dimensions.   

\end{abstract}
\section{Introduction}
d=5 N=2 U(1) gauged Supergravity is interesting as it is the simplest supergravity theory in 5 dimensions $\cite{gunaydin}$ and it can be shown to arise from compactification of M-theory on a Calabi-Yau 3-fold $\cite{chou}$.  It has also enjoyed a ressurgence of interest recently as various authors (e.g. $\cite{kalinde}$,$\cite{cvetic}$) have attempted to find a M-theoretic realisation of the Randall-Sundrum models $\cite{rs}$.  The bosonic part of the un-gauged Lagrangian can be written

\begin{equation}
e^{-1}{\cal L}=-\frac{1}{2}R-\frac{1}{4}G_{IJ}(\phi)F_{\mu\nu}^{I}F^{\mu\nu{J}}-\frac{1}{2}g_{\mu\nu}\bar{g}_{xy}(\phi)\partial^{\mu}\phi^{x}\partial^{\nu}\phi^{y}+\frac{e^{1}}{48}\epsilon^{\mu\nu\rho\sigma\lambda}C_{IJK}F_{\mu\nu}^{I}F_{\rho\sigma}^{J}A_{\lambda}^{K}
\end{equation}

where $g_{\mu\nu}$ is the 5-D space-time Metric, $\mu,\nu$=0,..,4.  The couplings of the n scalars $\bar{g}_{xy}$, $x,y$=0,..,n-1, form the metric of an n dimensional hypersurface embedded in an n+1 dimensional space.  This n+1 dimensional space is parametrised by the coordinates $X^I$ and has a metric $G_{IJ}$, $I,J$=0,..,n, which also gives the vector couplings in the Lagrangian.  When considering the above lagrangian in the context of M-theory compactification, the total volume scalar of the Calabi-Yau is part of the universal hypermultiplet and, as here we only consider the vector multiplets, we are free to set it equal to 1.  The remaining degrees of freedom of the manifold are the K\"ahler moduli $t^{I}$ and they are related to the $X^{I}$ as follows $\cite{cham}$
\begin{equation}
V=C_{IJK}t^{I}t^{J}t^{K}=\frac{1}{6}C_{IJK}X^{I}X^{J}X^{K}|_{V=1}=1
\label{cubic}
\end{equation}
setting $X^{I}=6^{2/3}t^{I}|_{V=1}$.  In this context the $C_{IJK}$ are the intersection numbers of the Calabi-Yau, $C_{IJK}$ being constant and symmetric. The dimension of $t^{I}$ (and $X^{I}$) is equal to the number of harmonic (1,1) cycles of the manifold, so $n=h_{(1,1)}-1$.

The solution of this equation determines the n-dimensional scalar hypersurface and the metric on the (n+1)-dimensional space is given by
\begin{eqnarray}
G_{IJ}&=&-{\frac{1}{2}}{\frac{\partial^{2}}{\partial_{I}\partial_{J}}}(lnV)|_{V=1}{\nonumber}\\
&=&-\frac{1}{2}\left[C_{IJK}X^{K}-\frac{1}{4}(C_{IKL}X^{K}X^{L})(C_{JMN}X^{M}X^{N})\right]
\end{eqnarray} 
Usually we are free to identify the hypersurface $\phi^{y}$ directions with the embedding space $X^{y}$ directions allowing the metric on the scalar manifold to be written
\begin{equation}
\bar{g}_{xy}=G_{IJ}\frac{\partial{X^{I}}}{\partial{{\phi}^{x}}}\frac{\partial{X^{J}}}{\partial{{\phi}^{y}}}=G_{xy}+G_{y0}\frac{\partial{X^{0}}}{\partial{X^{x}}}+G_{x0}\frac{\partial{X^{0}}}{\partial{X^{y}}}+G_{00}\frac{\partial{X^{0}}}{\partial{X^{x}}}\frac{\partial{X^{0}}}{\partial{X^{y}}}.
\end{equation}

The gauging of the theory is achieved by turning on the M-theory 4-form field strength in the internal directions of the Calabi-Yau which gives a constant $n+1=h_{(1,1)}$ dimensional vector field, $\alpha_{I}$, throughout the (n+1)-dimensional embedding space $\cite{stelle}$.  This yields the superpotential 
\begin{equation}
W=\alpha_{I}X^{I}
\label{w}
\end{equation}
and the scalars obtain a potential
\begin{equation}
U=6\left(\frac{3}{4}\bar{g}^{xy}\frac{\partial{W}}{\partial{\phi^{x}}}\frac{\partial{W}}{\partial{\phi^{y}}}-W^{2}\right).
\end{equation}

The setup is now completely determined by $C_{IJK}$ and $\alpha_{I}$, but there are some constraints on the kind of supergravities allowed.  Firstly, supersymmetry requires $\bar{g}_{xy}$ to be such that the embedding space metric $G_{IJ}$ must be positive definite $\cite{gun2}$.  Secondly, to avoid unphysical situations, the kinetic term for the scalars must obey the constraint $g_{\mu\nu}\bar{g}_{xy}\partial^{\mu}\phi^{x}\partial^{\nu}\phi^{y}\ge 0$.  The positivity of the vector coupling places restraints on the intersection matrix.

\section{Multiple Branches and Incomplete Spaces}

To visualise some of the features of this situation, one can consider the configuration examined in $\cite{gun2}$.  This corresponds to the intersection matrix
\begin{equation}
C_{000}=1, \qquad C_{0xy}=-\frac{1}{2}\delta_{xy}, \qquad C_{00x}=0
\label{constraint}
\end{equation}

\begin{figure}[tb]
\centering
\includegraphics[width=10.5cm,height=8cm]{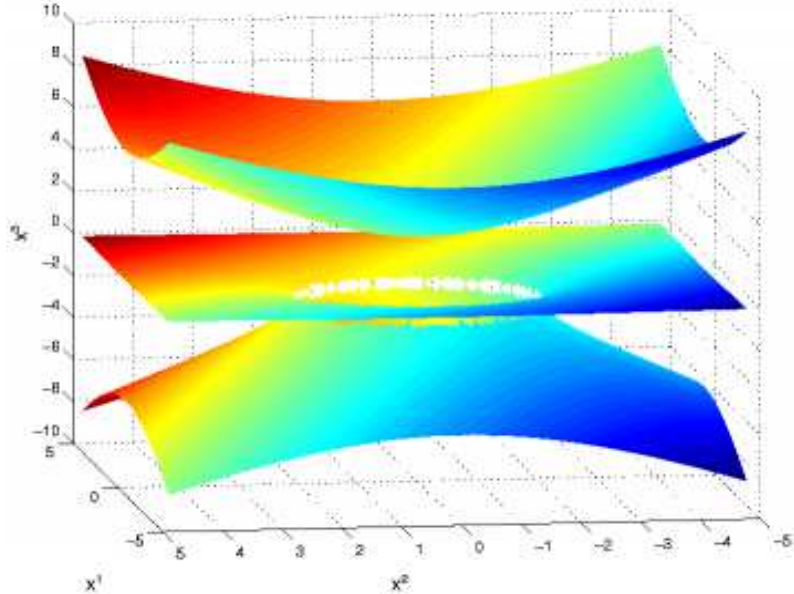}
\caption{Scalar manifold embedded in (n+1)-dimensional space for $C_{IJK}$ given in the text}
\label{whole}
\end{figure}
The scalar manifold corresponding to these intersection numbers has been plotted in figure $\ref{whole}$ and consists of three surfaces, two of which meet at the circle where the roots defining them start to become complex conjugate.  The shading on the plot shows the values of superpotential for a gauging vector $\alpha_{0}=0, \alpha_{x}=1$.  In this diagram, it is possible to see the non-trivial topologies that can be obtained where two solutions of the cubic equation meet.  

It is tempting to try and utlise these topological features to evaid the results of $\cite{kalinde}$ and $\cite{cvetic}$ which show the difficulties in obtaining a domain wall solution in this type of supergravity. However the internal metric of the scalar manifold is positive definite on the (upper) hypersurface which is everywhere real, but not on the other two surfaces.  There are several interesting topologies which can arise upon solution of the cubic polynomial for other intersection numbers but none of them immediately lend themselves to domain wall construction.

There are also matrices corresponding to situations where only part of a manifold has a positive definite scalar metric and vector coupling, whereas other regions are non positive definite.  Some of the allowed regions also posess interesting topologies, but before presenting an example, it is important to consider what happens as we pass from a positive region into a negative one. 

\begin{figure}
\begin{tabular}{ccc}
\includegraphics[height=4.5cm,width=7cm]{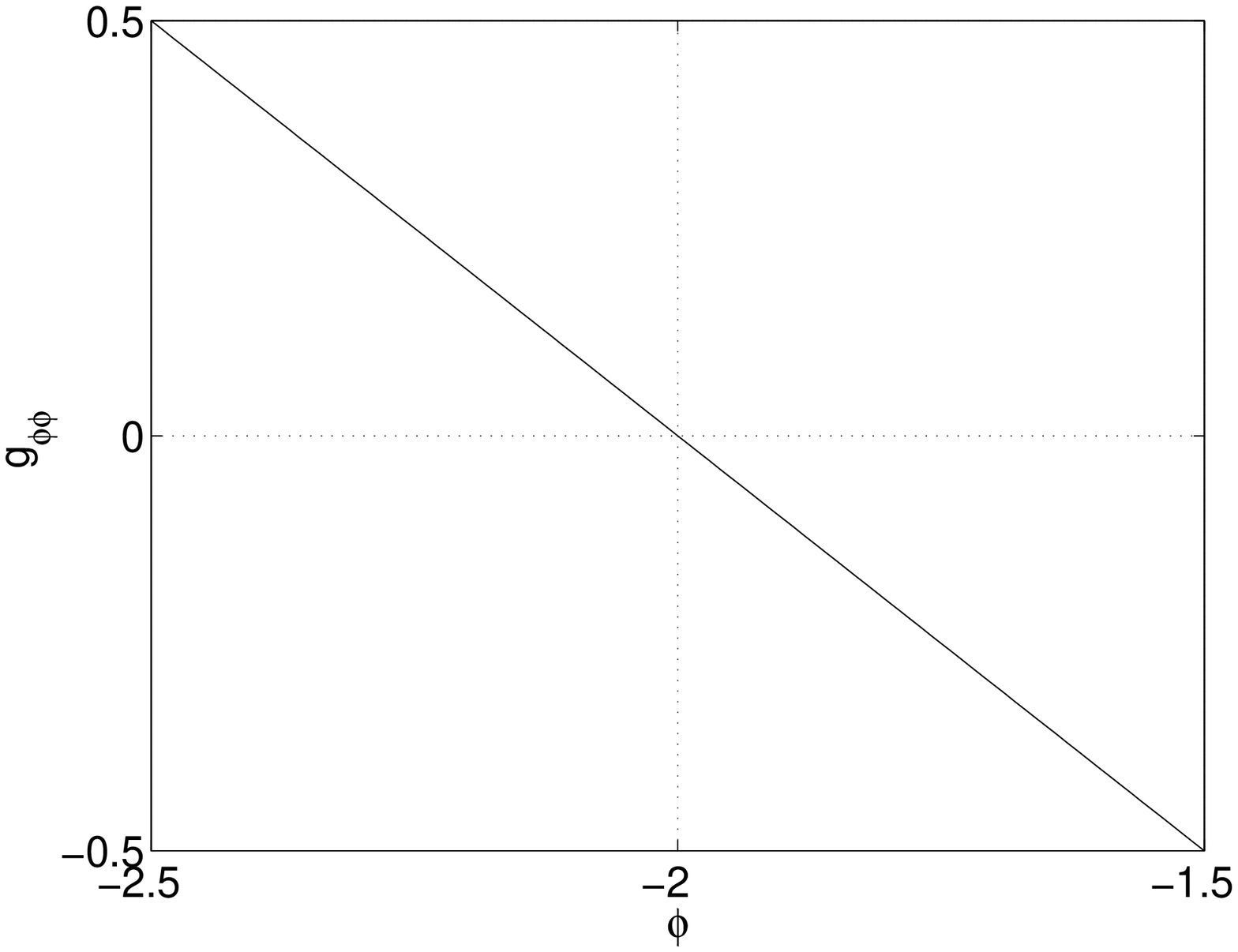} 
&\hspace{1.0cm}&
\includegraphics[height=4.5cm,width=7cm]{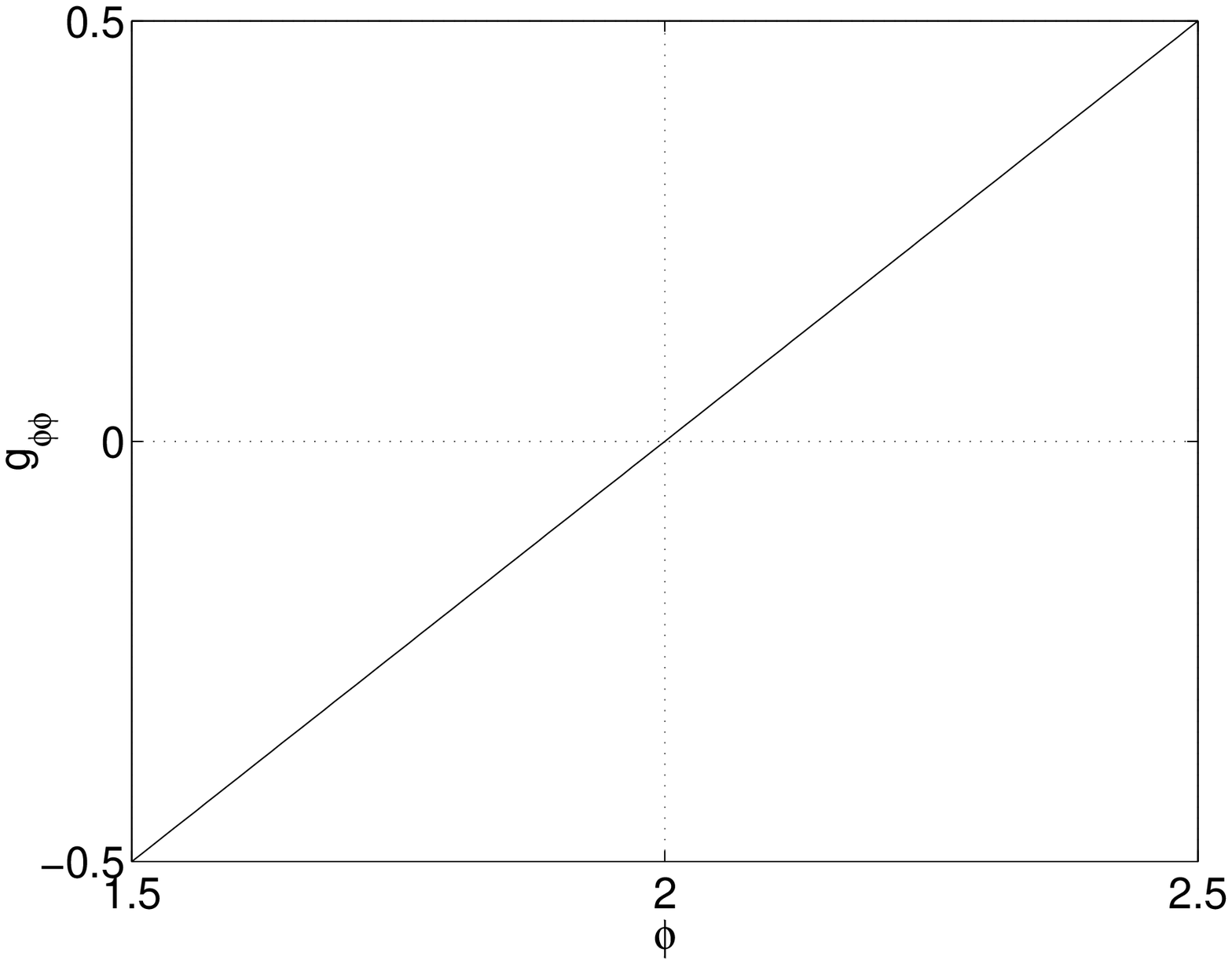}\cr
{\bf $g_{\phi\phi}$} & \hspace{1.0cm}& {\bf $g_{\phi\phi}$} \cr
\includegraphics[height=4.5cm,width=7cm]{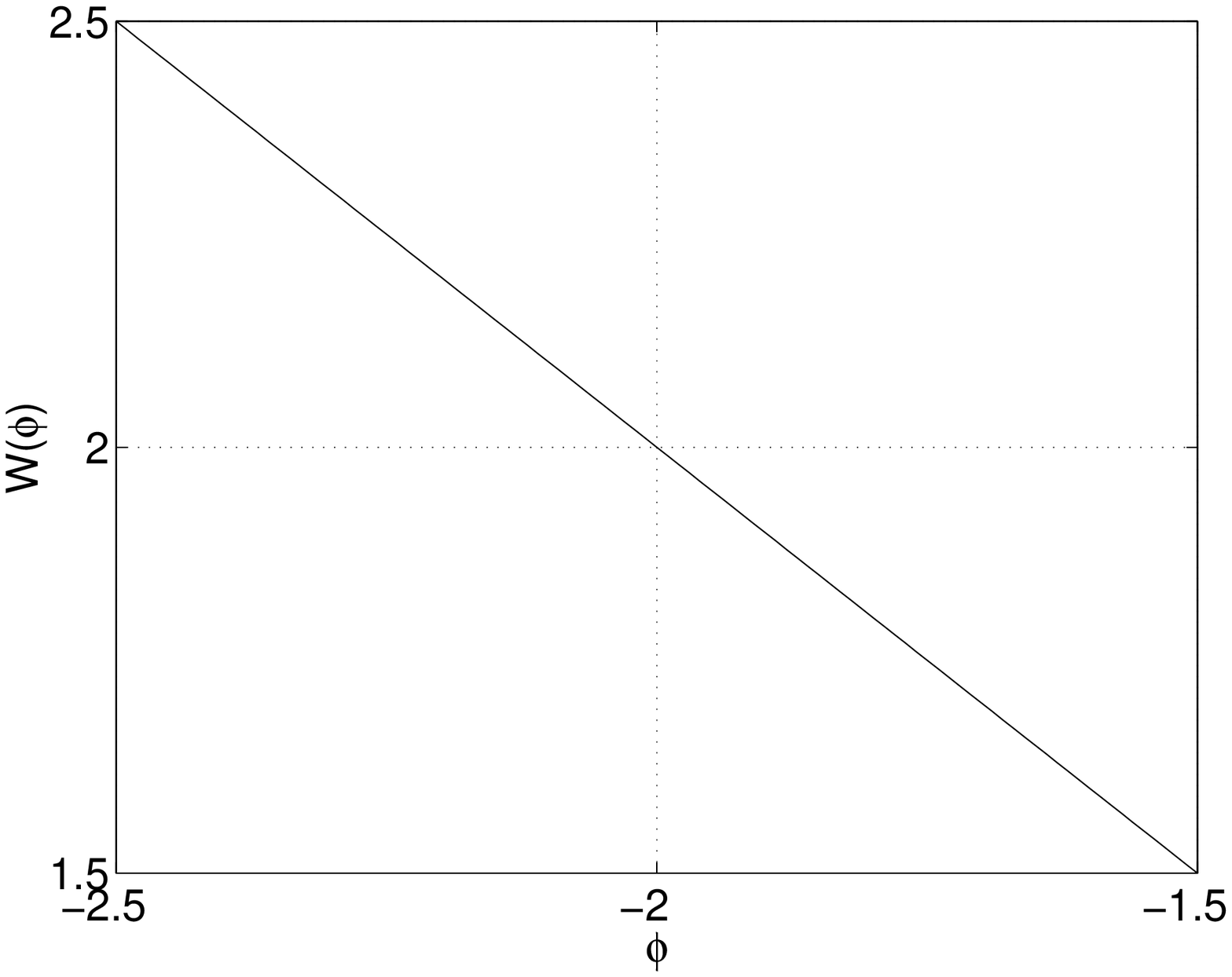}
&\hspace{1.0cm}&
\includegraphics[height=4.5cm,width=7cm]{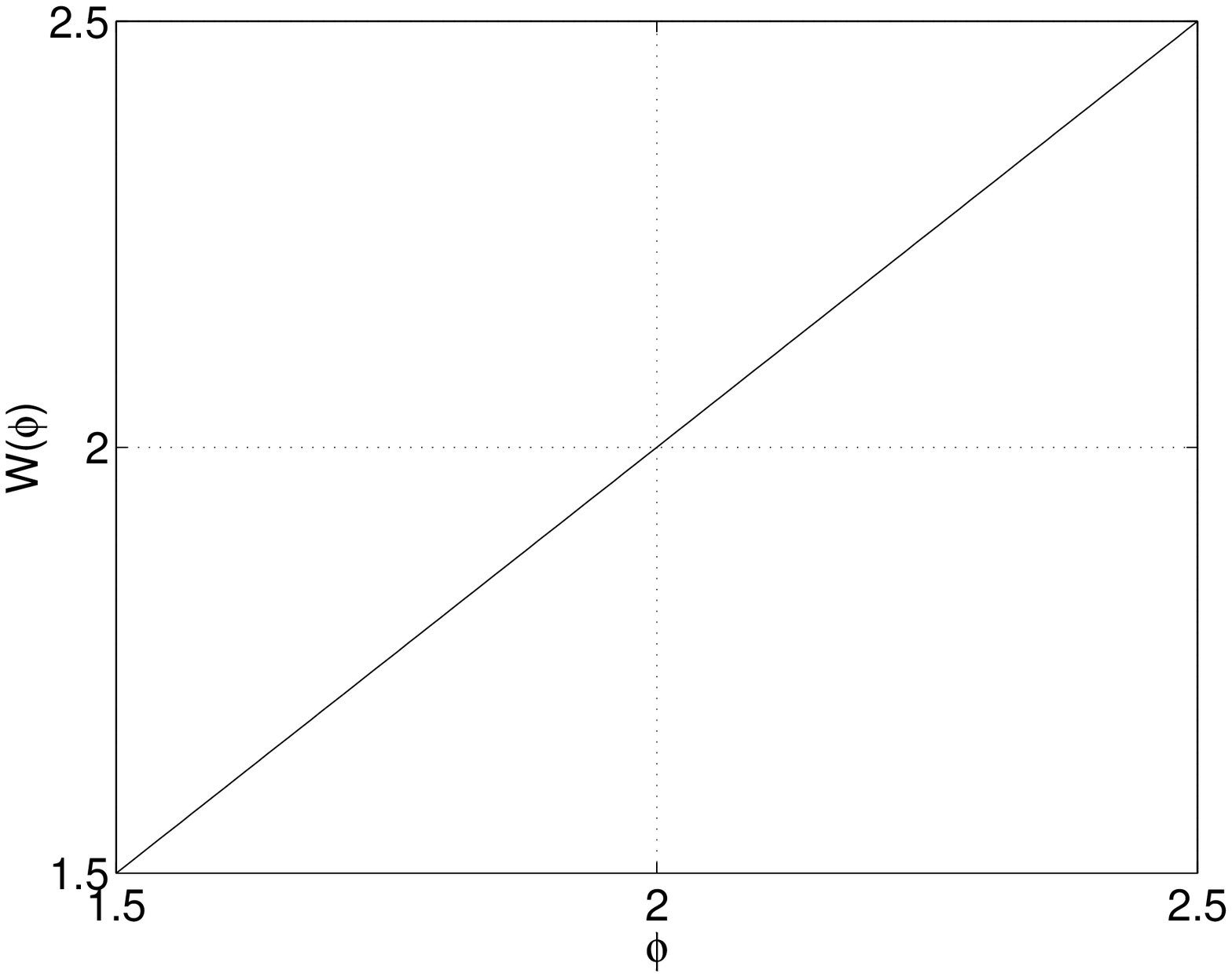}\cr
{\bf $W(\phi)$} & \hspace{1.0cm}& {\bf $W(\phi)$} \cr
\includegraphics[height=4.5cm,width=7cm]{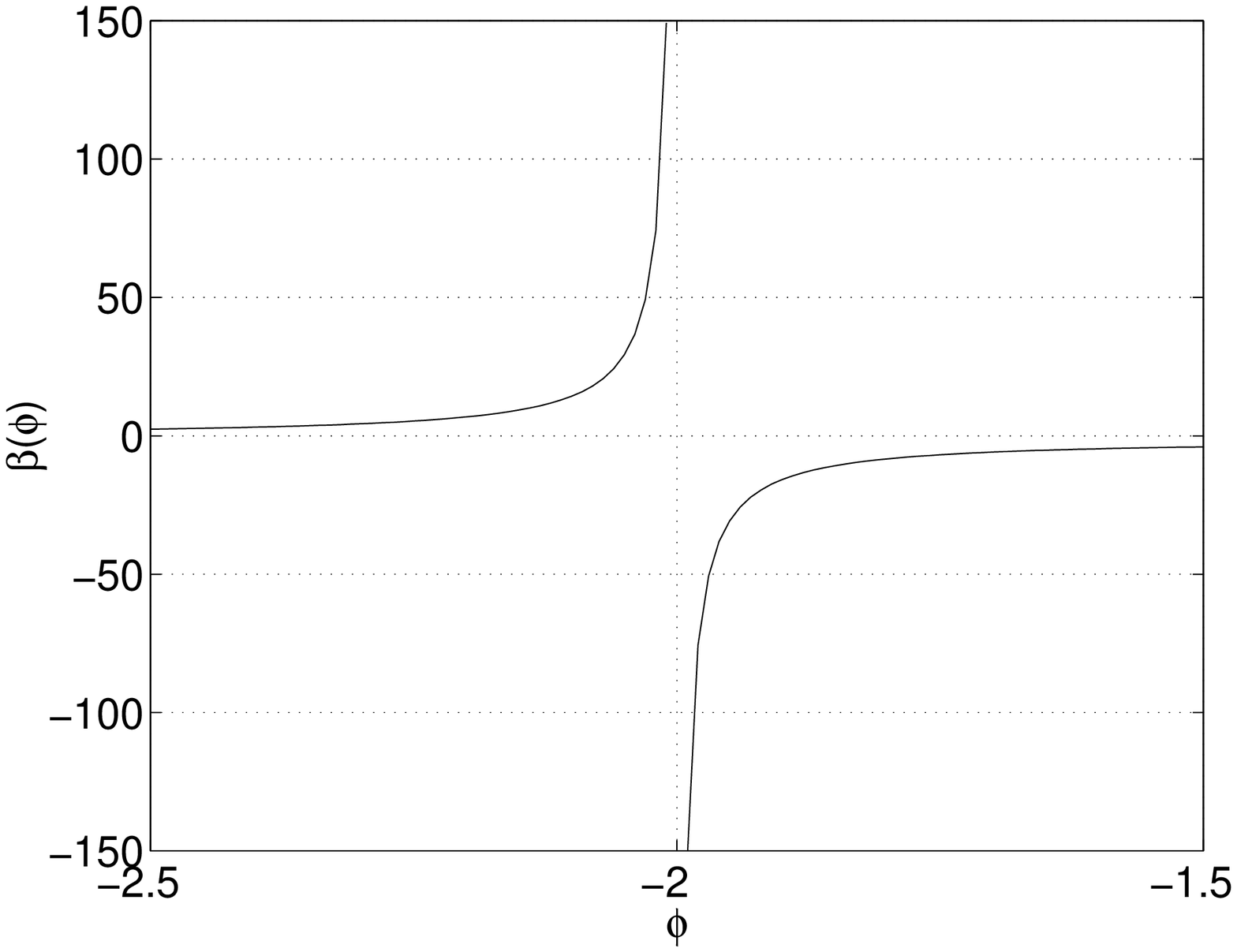}
&\hspace{1.0cm}&
\includegraphics[height=4.5cm,width=7cm]{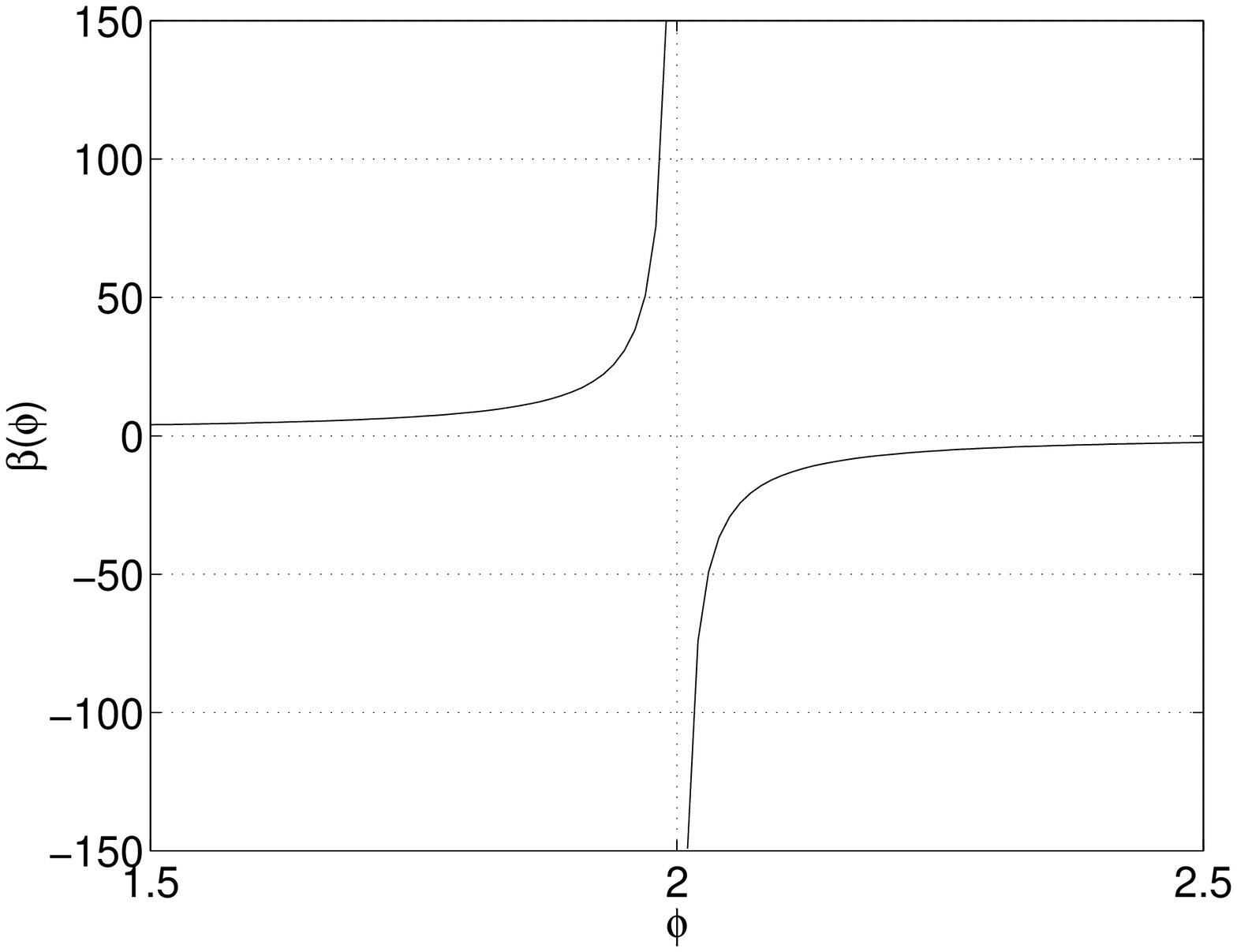}\cr
{\bf $\beta(\phi)$} & \hspace{1.0cm}& {\bf $\beta(\phi)$} \cr
\end{tabular}    
\caption{Toy model showing the behaviour of the beta function as the scalar metric changes sign.  $\beta(\phi)=-3g^{\phi\phi}\partial{W}/W$}
\label{toy}
\end{figure}

To do so we write the metric in the following way.
\begin{equation}
ds^{2}=\mu^{2}(-dt^{2}+dx_{3}^{2})+\frac{d\mu^{2}}{\mu^{2}W(\mu)^{2}}
\end{equation}
Now we can write the progress through moduli space as we move through the fifthe dimension as a beta function $\cite{behrndt2}$
\begin{equation}
\beta^{x}=\mu\frac{d}{d\mu}\phi^{x}=-3g^{xy}\frac{\partial_{y}W}{W}
\end{equation}
By investigating an idealised situation with one scalar $\phi$ and a superpotential and metric behaving in the way illustrated in figure $\ref{toy}$ we can deduce what will happen as the flow crosses between the two regions.  The flow is attracted to the point where the metric goes to zero, although it cannot cross that point, as if it were to cross onto the other side there would be an infinite repulsion. In this way the fields will be attracted to the edge of the region which posesses a positive definite manifold.

\section{An Example}
Consider the situation correpsonding to two scalar fields and the intersection matrix
\begin{equation}
C_{000}=1, \qquad C_{0xy}=-\delta_{xy}, \qquad C_{00x}=0, \qquad C_{222}=1, \qquad C_{221}=-1.
\label{matrix2}
\end{equation}
\begin{figure}[tb]
\centering
\includegraphics[width=10.5cm,height=8cm]{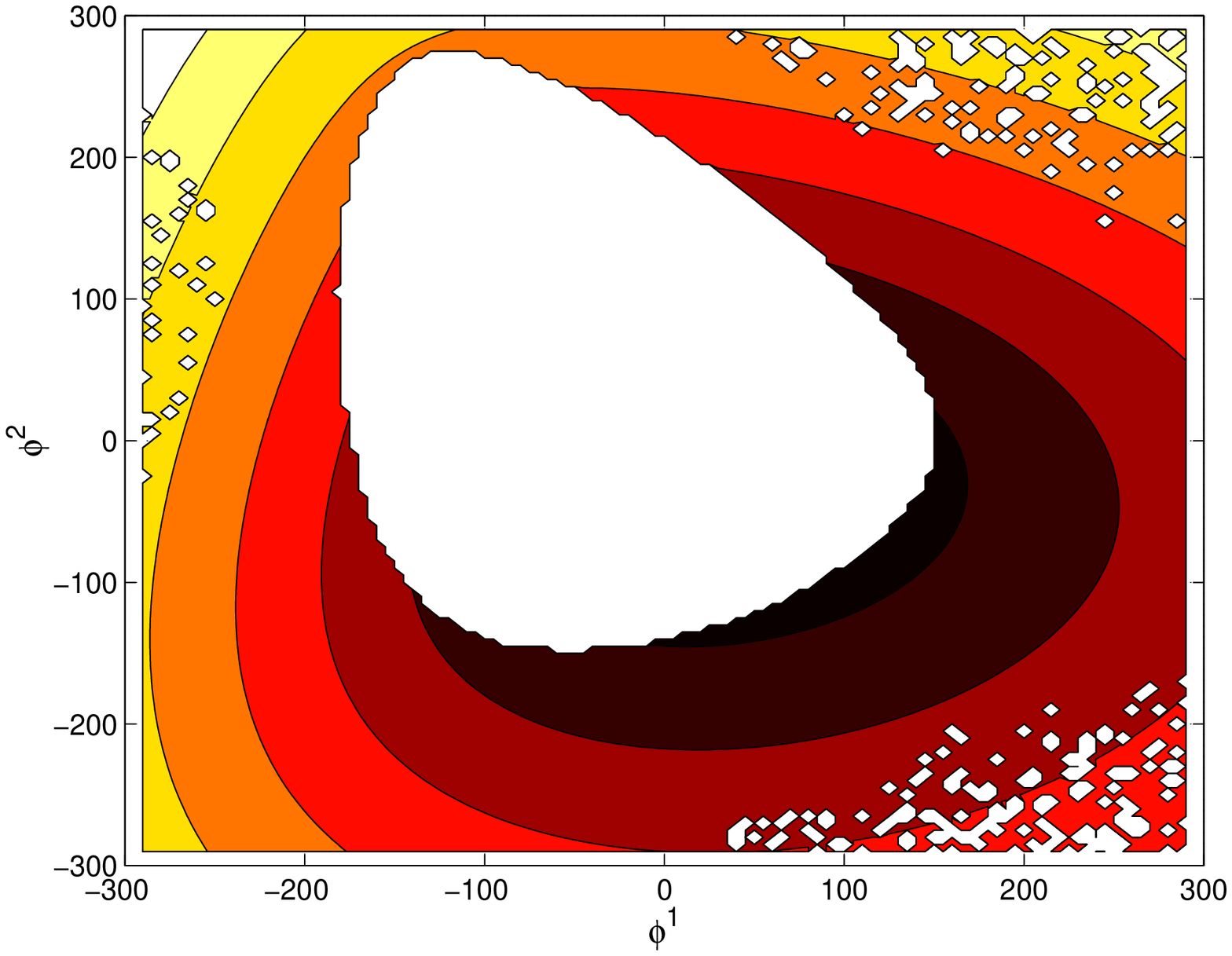}
\caption{Scalar manifold obtained from the $C_{IJK}$ in ($\ref{matrix2}$).  Only regions with positive scalar metric have been plotted and lighter shading corresponds to large superpotential}
\label{ring}
\end{figure}
If one takes the cubic surface corresponding to the solution of the polynomial obtained from these intersection numbers and then calculates which regions of this surface posess a positive definite scalar metric, the surface generated contains a central region with non-positive scalar metric into which the scalar field will not flow, although the flow will be attracted to the edge of this region. We are again free to choose the gauging vector $\alpha_{0}=1, \alpha_{1}=-1/2, \alpha_{2}=1/2$ and then plot the allowed region of this manifold and the superpotential within that region (see figure $\ref{ring}$).

There are no fixed points within the allowed region which satisfy the criterion $\frac{\partial{W}}{\partial\phi^{x}}=0$, but there are points on the edge of the disallowed central region which act as effective saddle points.  Normal to the dividing line the fields are trapped on one side by the increasing superpotential and on the other by the forbidden region.  Parallel to the dividing line the superpotential falls away in both directions.  Without such a mechanism it is impossible to obtain saddle points in the superpotential in this simplest of d=5 supergravities $\cite{cvetic}$. 

The choice of gauging vector was made to provide an offset superpotential with a minimum and a maximum provided the fields are trapped to the edge of the forbidden region.  The situation is reminiscent of an axion potential where the symmetry of a U(1) minimum is lifted by the addition of another potential slanted in one direction. This configuration is interesting as it forces the fields away from what would be the trivial situation of only one $AdS_{5}$ vacuum at the critical point of the superpotential.  This point would only be accessible if the central region of this solution had a positive definite scalar metric.  Also, because of the non trivial topology of the permitted region, field configurations may exist where the one minimum of the superpotential serves as both vacua of a domain wall solution, again reminiscent of axionic domain walls.

\section{Geometry of the Effective Fixed Points} 

Although this situation seems to solve some of the problems which were first clarified in $\cite{kalinde}$, it is still not a viable realisation of the Randall Sundrum models.  In these models, it is neccesary to have two $AdS_{5}$ minima on either side of an interpolating saddle point (assuming no imposition of $Z_{2}$ symmetry).  The criteria for this has been well studied, see e.g. $\cite{chou}$.  At the critical points where $AdS_{5}$ is desired, the normal to the scalar manifold must be parallel to the gauging vector $\alpha_{I}$.  Using the very special coordinates $t^{I}$ and following the conventions of $\cite{cham}$ the normal is written
\begin{equation}
\frac{\partial V}{\partial t^{I}}=\frac{\partial}{\partial t^{I}}C_{LMN}t^{L}t^{M}t^{N}=3t_{I}.
\end{equation}
As we are free to choose the gauging vector, it is possible to obtain the corrct vacua, although how contrived that choice is would ultimately depend upon the details of the compactification from 11 dimensions.  Finding a BPS solution around the loop is more difficult.  As pointed out in $\cite{behrndt2}$, the neccesary condition for a BPS configuration is that the normal to the hypersurface remains parallel to the harmonic function of the solution and this problem will be addressed in future work.

In this letter we have presented the possibility that the topology of the allowed regions on the scalar manifold, i.e. those regions with positive definite vector and scalar metrics, can in certain situations be non-trivial and this may aid the construction of domain walls solutions.  While this work was in progress a number of papers were published with alternative and interesting solutions to these problems (see $\cite{duff}$,$\cite{thomas}$\&$\cite{adam}$).  I have also been made aware of some relevant previous work e.g. $\cite{sabra}$.

\section*{Acknowledgements}
I benefited during the early part of this work from conversations with David Bailin, Klaus Behrndt and Tom Dent.  I also owe a debt of gratitude to the Sussex astronomy group for their help with FORTRAN, especially Joy Muanwong.


\begin{thebibliography}{99}
\bibitem{gunaydin}G\"{u}naydin, M., Sierra, G. and Townsend, P.K., Nucl.Phys.{\bf{B253}}, 573, (1985)
\bibitem{chou} Chou, A., Kallosh, R., Rahmfeld, J., Rey, S.-J., Shmakova, M. and Wong, W.K., Nucl.Phys.{\bf{B508}}, 147, (1997) 
\bibitem{kalinde} Kallosh, R. and Linde, A., JHEP {\bf{2}}, 5 (2000) 
\bibitem{cvetic} Behrndt, K. and Cvetic, M., Phys.Rev.{\bf{D61}}, 101901, (2000)
\bibitem{rs} Randall, L. and Sundrum, R., Phys.Rev.Lett.{\bf{83}}, 3370,(1999) and also Phys.Rev.Lett.{\bf{83}}, 4690, (1999)
\bibitem{cham} Chamseddine, A., Ferrara, S., Gibbons, G.W. and Kallosh, R., Phys.Rev.{\bf{D55}}, 3647, (1997)
\bibitem{stelle} Stelle, K.S., hep-th/9812086
\bibitem{gun2}G\"{u}naydin, M., Sierra, G. and Townsend, P.K., Class.Quantum.Grav.{\bf{3}}, 763, (1986)
\bibitem{gun3}G\"{u}naydin, M., Sierra, G. and Townsend, P.K., Nucl.Phys.{\bf{B242}}, 244, 1984
\bibitem{gun4}G\"{u}naydin, M. and Zagermann, M., hep-th/9912027
\bibitem{behrndt1} Behrndt, K. and Gukov, S., hep-th/0001082
\bibitem{behrndt2} Behrndt, K, Nucl.Phys.{\bf{B573}}, 127, 2000 
\bibitem{duff} Duff, M.J., Liu, J.T. and Stelle, K.S., hep-th/0009212 and Duff, M.J., Liu, J.T. and Sabra, W.A., hep-th/0007120
\bibitem{thomas}  Behrndt, K., Herrmann, C., Louis, J. and Thomas, S., hep-th/0008112
\bibitem{adam} Falkowski, A., Lalak, Z. and Pokorski, S., hep-th/0009167
\bibitem{sabra} Chamseddine, A.H. and Sabra, W., Phys.Lett. {\bf{B477}}, 329, (2000)
\end{thebibliography}
\end{document}